\documentstyle[prl,aps,epsf]{revtex} 

\draft

\def\Bsub{{\it Bacillus subtilis}} \def\bsub{{\it B. subtilis}}

\def\Ecol{{\it Escherichia coli}} \def\ecol{{\it E. coli}}

\def\Mjan{{\it Methanococcus janaschii}} \def\mjan{{\it M. janaschii}}

\def\olig{{oligonucleotide}}  \def\oligs{{\olig s}}
  
\def\kmer{{$k$-mer}}          \def\kmers{{\kmer s}}
     \def\sig{{$\sigma$}} 
    
\def\dist{{distribution}}  \def\dists{{\dist s}}

\def\rmd{{root-mean-deviation}}

\begin{document}

\twocolumn[\hsize\textwidth\columnwidth\hsize\csname %
@twocolumnfalse\endcsname
\phantom{xx}
\vskip -1.4cm
%\hfill{{\it Preprint} NCU/CCS-2002-0525; NSC-NCTS/Phys-020601}

\medskip

\title
{\bf Minimal model for genome evolution and growth}

\author{Li-Ching Hsieh$^\ast$, Liaofu Luo$^\dag$, Fengmin Ji$^\ddag$ 
and H.C. Lee$^{\ast\S}$}
\address{
$^\ast$Department of Physics,  
National Central University, Chungli, Taiwan 320\\
$^\dag$Department of Physics, University of Inner Mongolia
Hohhot 010021, China\\
$^\ddag$Department of Physics, Northern JiaoTong University,
Beijing 100044, China\\
$^\S$Department of Life Science,
National Central University, Chungli, Taiwan 320}

\date{\today}

\maketitle
\begin{abstract}
Textual analysis of typical microbial genomes reveals that they have
the statistical characteristics of a DNA sequence of a much shorter
length.  This peculiar property supports an evolutionary model in
which a genome evolves by random mutation but primarily grows by
random segmental self-copying.  That genomes grew mostly by
self-copying is consistent with the observation that repeat sequences
in all genomes are widespread and intragenomic and intergenomic
homologous genes are preponderance across all life forms.  The model
predicates the coexistence of the two competing modes of evolution: 
the gradual changes of classical Darwinism and the stochastic 
spurts envisioned in ``punctuated equilibrium''.
\end{abstract}

]

%\bn{\bf Introduction}.
The genome of any organism extant is the culmination of a long 
history of growth and evolution that extends back to the origin 
of life.   How much about this history can we learn from 
the present state of that genome? 
When a genome is viewed as a text composed of the four 
``letters'' A (adenine), C (cytosine), G (guanine) and T (thymine), 
it is essentially a random text.  This is so because, as far as we know, 
genomes are made by a ``blind watchmaker'' \cite{Dawkins}.  Whatever 
is not random about a genome is caused by the forces of selection 
that indirectly,  subtly and slightly favor some random 
patterns over others.  This is why it is such a challenge to delineate 
coding  parts of a genome including genes and regulatory sequences 
from noncoding parts, and especially so 
when the effort cannot benefit from sequence similarity to other 
known coding sequences \cite{Hogenesch}. 

%\sn {\bf Frequency of occurrence of oligonucleotides.}
Yet the randomness of genome is not of the trivial kind.    
An example that hints at the potential complexity of the 
genome-as-text is the distribution 
of the frequency of occurrence of \oligs.   
(In what follows, 
frequency will always mean frequency of occurrence, 
a \kmer\ is an \olig\ of length $k$ 
and a \dist\ of frequency of \kmers\ will be called a \kmer\ \dist.)
Frequency of short \kmers\ has been 
used in studies of molecular evolution \cite{Burge,Luo95,Karlin}. 
The frequency of a \kmer\ is the number of times it is seen through  
a sliding window of width $k$ when it traverses once across the genome. 
If the length of the genome is $L$, 
the act just described is similar to distributing $L$ objetcs (we think 
of the genome as being circular) into $4^k$ boxes, the total number 
of different \kmers.    Hence when $L$ is much greater than 
$4^k$, the \kmer\ \dist\ for a simple random genome 
sequence is expected to be a Poisson distribution 
with the mean and deviation both being $L/4^k$.  By a simple random 
genome sequence of a given base composition 
we mean the sequence that would obtain when any sequence of that 
base composition is thoroughly scramble.

\begin{figure}[tbh]
\begin{center}
\epsfxsize=8cm\epsfysize=4.5cm 
%\epsffile{/home/hclee/DNA/model/gro/Fig_ecol_ran.ps}
\epsffile{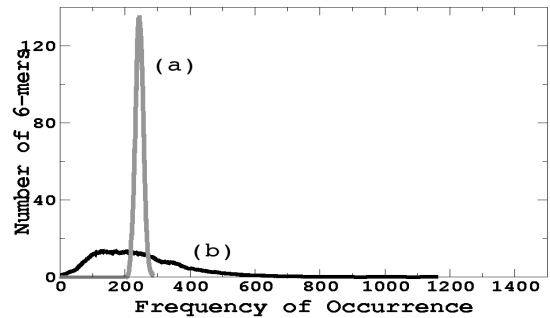}
\end{center}
\caption{\label{Ecol_ran} \baselineskip=6pt\small\sf 
Distribution of frequency of 6-mers of (a) a simple random 
sequence 1 Mb long with 50\% A+T content and (b) the genome of
\ecol, whose A+T content is approximately 50\%.}
\end{figure}

Figure \ref{Ecol_ran}(a)  
shows the 6-mer \dist\ in a simple random 
sequence of length one million bases (1 Mb) with unbiased
base composition.  The mean of 244 and \rmd\ of 15.5 
characterize the \dist\ as being Poisson. 
Figure \ref{Ecol_ran}(b) is the distribution obtained from the 
complete genome of \Ecol\ \cite{ecol} 
whose base composition is essentially unbiased.  
(Microbial complete genome sequences are taken from the 
GenBank \cite{GenBank}.  In this work, the frequencies of \kmers\ in 
microbial complete genomes are normalized to correspond to those 
of a 1 Mb long sequence by multiplying each frequency by a factor 
equal to 10$^6$ divided by the length of the genome.)  
While strikingly different from Fig. \ref{Ecol_ran}(a),  
Figure \ref{Ecol_ran}(b) is 
representative of microbial complete genomes with 
an unbiased base composition.  It has a \rmd\ (140) that is nine 
times that of the simple random sequence.  Whereas simple random sequence 
contains no 6-mers whose frequency is greater than 
400 or less than 100, the corresponding numbers of 6-mers in the genome of 
\ecol\ are about 500 and 510, respectively.

\begin{figure}[tbh]
\begin{center}
\epsfxsize=8cm\epsfysize=4.5cm 
%\epsffile{/home/hclee/DNA/model/gro/Fig_mjan_ran.ps}
\epsffile{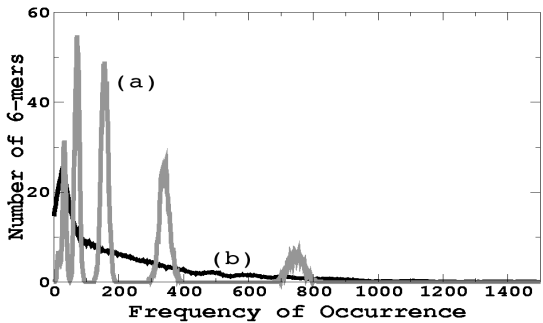}
\end{center}
\caption{\label{Mjan_ran} \baselineskip=6pt\small\sf 
Distribution of frequency of 6-mers of (a) a simple random 
sequence 1 Mb long with 70\% A+T content and (b) the genome of
\mjan, whose A+T content is approximately 70\%.   The positions of 
the peaks in (a) are explained in the text.}
\end{figure}

Figure \ref{Mjan_ran}(a) shows the distribution of a simple random 
sequence whose A+T content is 70\% (it is a general fact 
of genomes that the number of A and T bases are almost the same, 
similarly for C and G contents).  
The single narrow peak seen in Figure \ref{Ecol_ran}(a) is 
now broken into seven smaller peaks whose appearance is 
caused by the bias in the base composition; 
the mean frequency of 6-mers with $m$ A or T's is
244$\times(7/5)^m (3/5)^{6-m}$, giving the positions of the 
seven peaks to be  
11.4, 26.6, 62.0, 144, 337, 787, 1837, for $m=$ 0 to 6, 
respectively (the last peak is off scale in  Figure \ref{Mjan_ran}). 
%The \rmd\ of the \dist\ is 264, a large value reflecting  
%the highly biased base composition.  
Figure \ref{Mjan_ran}(b) is the distribution obtained 
from the complete genome of
\Mjan\ whose A+T content is approximately 70\% \cite{mjan}.  
Although both \dists\ are very broad and have large values for 
their \rmd s - 264 for (a) and 320 for (b) - that reflect an underlying 
highly biased base composition,  that for the genome of \mjan\ is 
significantly greater and the two \dists\ are in any case 
again clearly dissimilar in detail.

When the  7-, 8- and 9-mer \dists\  
are examined, the discrepancy between a random sequence and complete 
microbial genomes persists.  We know of no previous explanation 
of this discrepancy.  Even as one would be  
tempted to attribute the cause of the discrepancy to biological effects, 
we shall show that that would likely be a wrong conclusion and that  
more likely the observed 
\dists\ have an interesting stochastic origin. 
(When $k\ge 10$ the number of \kmers\ becomes too large 
for the Poisson distribution to be a reliable yardstick for 
judging whether a genome is simply random or not.)

Although genomes are of the order of 1 Mb long, the ratio of the mean of the 
6-mer \dist\ to its \rmd\ suggests the statistical property of 
a much shorter sequence, perhaps as short as 10 kb.    
In \ecol, that ratio is 1.74 (as opposed to a ratio of $\sqrt{244}$=15.6 
for a simple random sequence 1 Mb long).  
Over all the complete microbial genomes, the \rmd\ after  
bias in base composition is corrected for 
ranges from 96 to 218 and has an average of 154.  This gives an average 
ratio of mean to \rmd\ of 1.58.    
In terms of a Poisson \dist\ such a 
ratio corresponds to a mean of 2.5 and a simple random sequence 
about 10 kb long, since there are 4096 6-mers. 
The 6-mer \dist\ of a 10 kb simple random sequence would have 
about 3310 of the 6-mers occur one to four times, 
3 to 4 of the 6-mers occur nine times and about one 6-mer occur 10 times. 
It would also have 350 of the 6-mers not occurring altogether.  
Suppose we now duplicate this simple random sequence 100 times to produce a 
1 Mb long sequence and let it undergo a number of single base 
mutations, then we may expect the long sequence to have a 6-mer \dist\ 
that begins to resemble Figure \ref{Mjan_ran}(b).   
That is, it should have many 
6-mers occurring more than 400 many times, some occurring close to 
1000 times, and many occurring fewer than 100 times.  

It may not be very realistic to suggest that real genomes 
are approximately high multiples of a much shorter sequence, plus 
mutation.  Sometime ago Ohno conjectured that great leaps 
in evolution had been the result of whole genome duplications \cite{Ohno}. 
The idea has remained controversial; the present state of gene sequence
information from vertebrates makes it difficult to either
prove or disprove this hypothesis \cite{Wolfe}, and phylogenetic studies 
of families of mammalian genes indicate that if ancient events of genome
duplication did occur, they did not play an important role in 
structuring the mammalian chromosomes bearing such genes \cite{Hughes}.  
In any case, even if events of whole genome duplication had occurred, 
it probably did not occur a very large number of times.  
On the other hand, there certainly have been a very large 
number of events of duplications of shorter sequences.   

Indeed most genomes have repetitive 
sequences (or repeat sequences) with lengths ranging 
from 1 base to many kbs whose numbers of 
copies far exceed those would be found in a simple random sequence. 
For example, in the  
human genome repeat sequences account for at least 50\% and
probably much more \cite{IHGSC,Celera}, because most ancient 
repeats presumably have been rendered unrecognizable as such by 
degenerative mutation.   
 
Here, we propose a minimal model for microbial genome growth 
that incorporates duplication of DNA of all lengths and that exhibits 
the observed \kmer\ distributions of real genomes.   The model 
employs the two types of events that drives genomic 
changes, mutation and DNA duplication.  For simplicity mutation 
events are represented by single base replacement (SBR).  
DNA duplication events are represented by occasional random duplication 
(RD) of a stretch of \olig\ with a 
characteristic length of \sig\ bases.   

In the model genomes are single stranded and 
the initial state of a genome is a simple random 
sequence of legth $L_0$ with a given base composition.  
From the initial state the genome evolves and grows by 
(base composition preserving) SBR and RD 
events until its length just exceeds 1 Mb.    
In an RD event, the length $l$ of the copied sequence is first 
randomly chosen (see below), then a site $p$ at least $l$ sites from 
the end of the genome is 
randomly chosen and the sequence from $p$ to $p+l-1$ is copied 
and inserted into the genome behind a second randomly chosen site. 
The model has three parameters: the initial length $L_0$, 
the ratio $\eta$ of the chances of having an SBR or an RD event 
and the length scale \sig. 
For the work reported here $L_0$ was held fixed at 1000 and 
only the two parameters $\eta$ and \sig\ were varied. 

At each instance of an RD event, a length $l$ not greater than 
the current length $L_c$ of the (artificial) genome for the duplicated 
segment is chosen as follows. We construct a function $G$ such that,
given a random number $y$ between zero and one, the
duplicated segment length is $l=G(\sigma;y)$.  
Let $w(x)$,  the probability per unit length 
of selecting a segment of length $x$, be proportional to  
$e^{-x/\sigma}$.  Then from $\int_0^{L_c} w(x) dx = 1$  one has  
$w(x) = \sigma^{-1} e^{-x/\sigma} (1- e^{-L_c/\sigma})^{-1}$. 
The recognition that inverse of $G$ is given by 
%\begin{equation}
$G^{-1}(l) = y = \int_0^l w(x) dx$ 
%= {1- e^{-l/\sigma}\over 1- e^{-L_c/\sigma}}
%\end{equation}
yields
\begin{equation}
l=G(\sigma;y) = -\sigma \ln[ 1-y(1-e^{-L_c/\sigma})]
\end{equation}
\noindent Note that when \sig$>>L_c$ the simplification $l \approx yL_c$ 
obtains.  When \sig$<<L_c$, $l \approx y\sigma$ when $y$ 
is close to zero, otherwise $1-y \approx e^{-l/\sigma}$ as long as 
$y$ is much greater than $e^{-L_c/\sigma}$ away from 1.   In all cases 
$G(1)=L_c$.  For fixed $L_c$ 
the average length of copied segments is 
$\bar{l} = \sigma - L_c e^{-L_c/\sigma}/(1-e^{-L_c/\sigma})$, which 
approaches $ \sigma$ when $L_c$ becomes much greater then \sig. 

Suppose the final genome length $L$ is much greater than $L_0$ and 
\sig\ (this will be the case here), then the total number of RD events 
will be somewhat greater than $L/\sigma$ and the total number of 
SDR events  will be somewhat greater than $\eta L/\sigma$. 

\begin{figure}[tbh] 
\begin{center}
\epsfxsize=8cm\epsfysize=4.5cm 
\epsffile{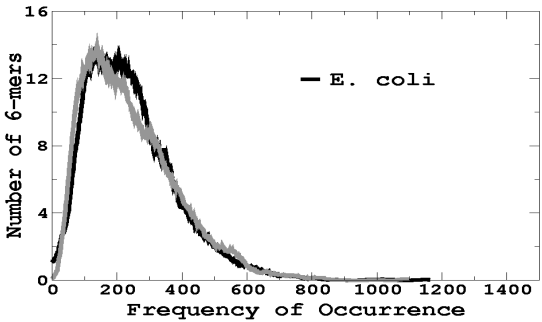}
\end{center}
%\par
\caption{\label{Ecol_rpt} \small\sf 
6-mer \dist\ of the genome of
\ecol\ (50\% A+T content) (black)
and a simple random sequence (50\% A+T content) 
including segmental duplication 
mechanism with $\eta = 500$ and $\sigma=15,000$ ( gray). }
\end{figure}

It turns out that if the model sequence is to have a 6-mer \dist\ 
similar to those of the representative 
real microbial genomes, the total number of mutations (for a sequence of 
canonical length 1 Mb) acting on the model sequence needs to be around 
40,000.  From the discussion in the previous section, this implies 
the relation $\sigma \approx 25\eta$ should hold.  The best results 
are obtained when $\sigma\approx 15,000$.  In Figure 3  
the model genome with an unbiased base composition generated 
with the parameters $\eta = 500$ and $\sigma=15,000$ is seen to 
have  a  6-mer \dist\ (gray) surprisingly similar to that of \ecol\ (black). 
No attempts were made to fine-tune the two parameters to get a 
``perfect'' fit. 
In Figure 4 (5, respectively) the distributions for the model genome 
(gray) generated with $\eta = 600$ and $\sigma=15,000$ and for 
the genome (black) of \Bsub\ \cite{bsub} (\mjan) are compared; 
both have approximately 60\% (70\%) A+T content.  
The peaks caused by biased base composition that one 
expects to see in a Poisson distribution (and seen in Figure 2(a)) are 
no longer evident in the \dists\ from the model genomes in Figures 4 and 5, 
just as they do not show in the \dists\ from real genomes.  
In particular, the model seems to succeed with ease in accounting 
for the very large number of 6-mers 
that occur with exceptionally high and with 
exceptionally low frequencies seen in most microbial genomes. 

\begin{figure}[tbh] 
\begin{center}
\epsfxsize=8cm\epsfysize=4.5cm 
%\epsffile{/home/hclee/talks/02/PS/Bsub_rpt.ps}
\epsffile{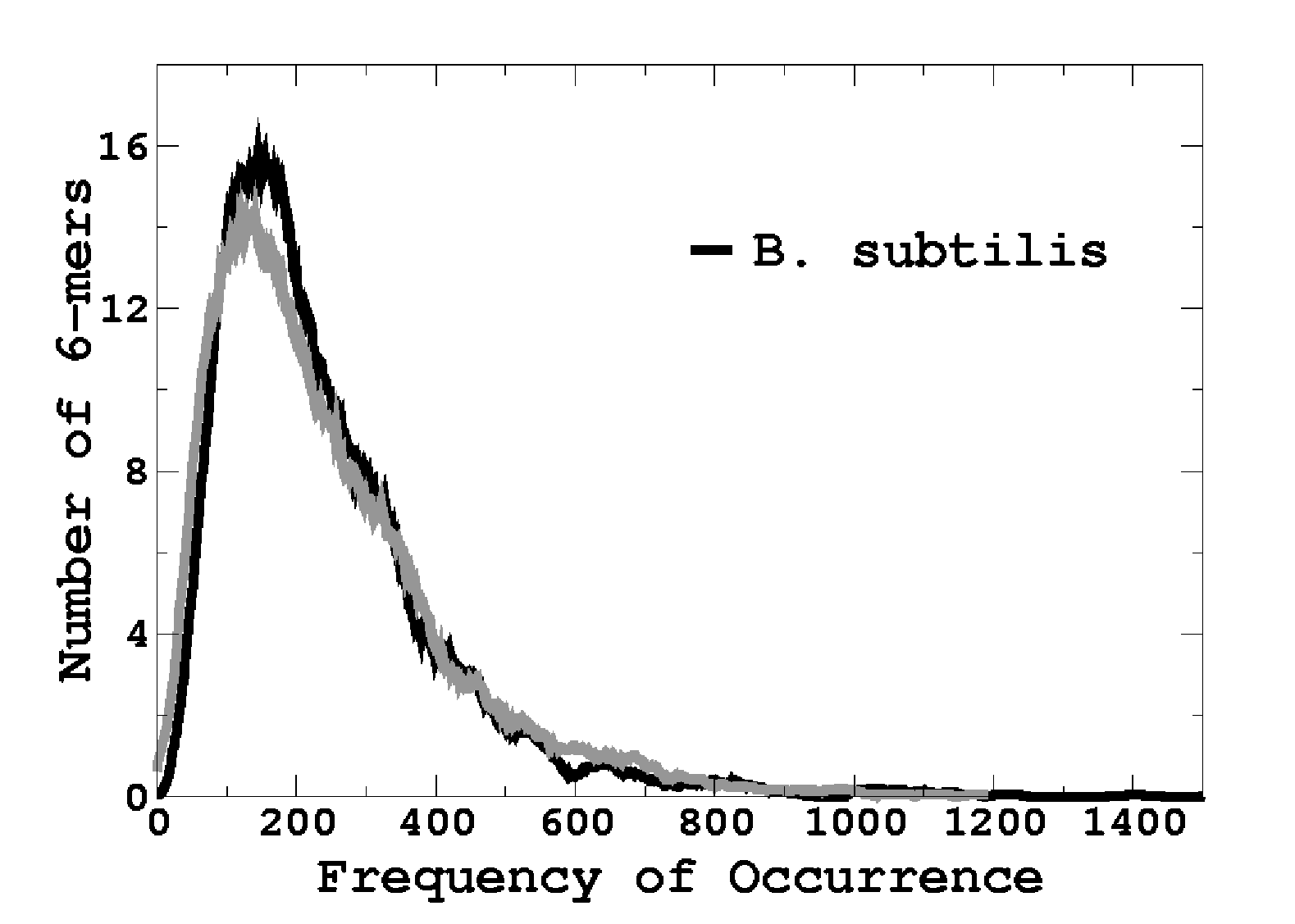}
\end{center}
%\par
\caption{\label{Bsub_rpt} \small\sf 
Same as Fig.\ref{Ecol_rpt}. Black:
\bsub\ (60\% A+T content); Gray: 
model sequence  with $\eta = 600$ and $\sigma=15,000$.}
\vskip-10pt
%\end{figure}
%\begin{figure}
\begin{center}
\epsfxsize=8cm\epsfysize=4.5cm 
\epsffile{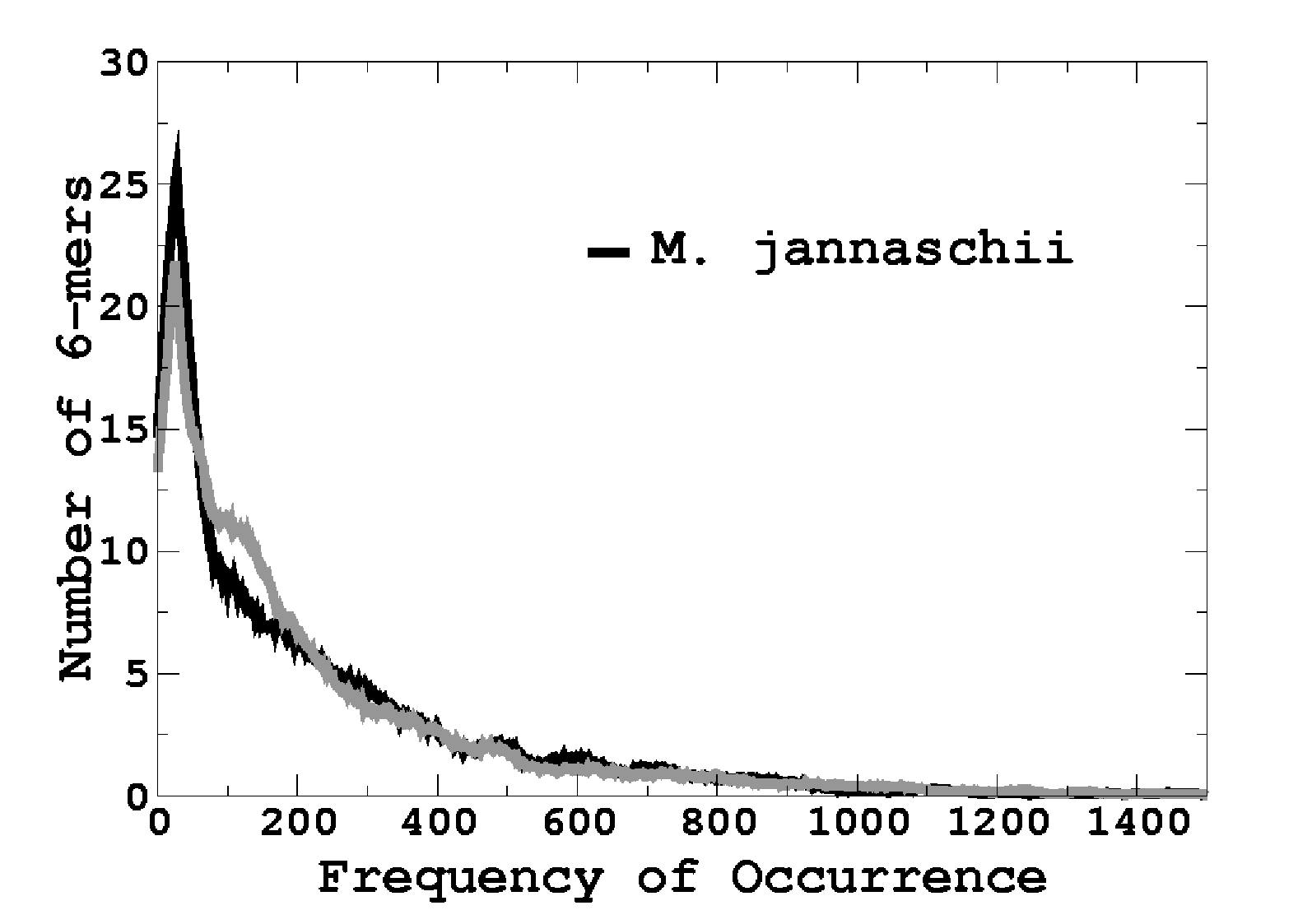}
\end{center}
%\par
\caption{\label{Mjan_rpt} \small\sf 
Same as Fig.\ref{Ecol_rpt}. Black:  
\mjan\ (70\% A+T content); Gray: 
model sequence  with $\eta = 600$ and $\sigma=15,000$.}
\end{figure}

It is emphasized that the high degree of likeness between the \dists\
of the simulated and real genomes notwithstanding, 
no claim is made of the general similarity of the contents of the 
two genomes.  If an alignment were to be made between, say, an 1 kb 
segment from the simulated genome and any segment of like length from 
the real genome, then the degree of similarity between them would be 
characteristic of that between two unrelated simple random sequences. 

The 6-mer \dists\ of microbial genomes are well represented by 
the two-parameter gamma \dist:
\begin{equation}
D(y) = y^{\alpha-1} \beta^{-\alpha} e^{-y/\beta}/\Gamma(\alpha)
\label{gamma} 
\end{equation}
The \dist\ has mean $\langle y\rangle= \alpha\beta$ 
and mean-square deviation 
$\Delta = \alpha^{1/2}\beta$.  In Table \ref{t:moments} the 
$n^{th}$ order deviations, defined as 
$\Delta^{(n)} = (\langle(y - \langle y\rangle)^n\rangle)^{1/n}$,
$n$ from 2 to 5,  of 6-mer \dists\ of real genomes are 
compared with those of: 
(a) the gamma distribution with the parameters $\alpha$ and 
$\beta$ (in brackets) obtained from the real genome distribution; 
(b) the 6-mer \dist\ of a simple random sequence without duplication; 
(c) the 6-mer \dist\ of the corresponding sequence given by the 
minimal model shown in Figures 3-5. 
The values of $\Delta^{(n)}$ in rows (a) show that the 6-mer distributions 
of the real genomes are well represented by gamma \dists.
The values of $\Delta^{(n)}$ in rows (c) show that 
the 6-mer distributions from the real and model genomes agree 
to a very high degree. 

\vbox{
\begin{table}[b]
\caption{\label{t:moments}
\small\sf High order deviations $\Delta^{(n)}$ 
of 6-mer \dists\ of microbial genomes and simple random sequences; 
$\Delta^{(2)}$ is the \rmd.  See text for description of 
the deviations and artificial sequences labeled (a), (b) and (c).} 
\begin{tabular}[t]{lcccc}
Sequence   & $\Delta^{(2)}$
& $\Delta^{(3)}$ & $\Delta^{(4)}$ & $\Delta^{(4)}$\\
\hline
 \ecol\ (50\% A+T content) & 140& 147& 213& 252\\
      (a) ($\alpha$ = 3.05, $\beta$ = 80.0)& 140 & 146 & 208 & 243\\ 
       (b) & 15.6 & 3.6 & 20.7 & 10\\ 
  (c) ($\eta$ = 500, $\sigma$ = 15K) & 144 & 148 & 212 & 247\\
\hline
      \bsub\ (60\% A+T) & 168 & 223 & 316 & 400\\ 
(a)  ($\alpha$ = 2.12, $\beta$ = 115)& 168 & 186 & 261 & 310\\ 
(b)  & 79 & 68 & 109 & 117\\
 (c) ($\eta$ = 600, $\sigma$ = 15K)& 169 & 194 & 266 & 311\\
\hline
      \mjan\ (70\% A+T) & 320 & 465 & 650 & 810\\ 
      (a) ($\alpha$ = 0.58, $\beta$ = 418)& 320 & 439 & 609 & 767\\ 
        (b) & 264 & 369 & 500 & 603\\ 
   (c) ($\eta$ = 600, $\sigma$ = 15K)& 321 & 462 & 635 & 783\\
\end{tabular}
\end{table}
}

A conspicuous deviation of the \kmer\ \dist\ of a real 
microbial genome from that of a simple random sequence is 
the very large numbers 
of extremely frequent and extremely rare 6-mers (or 7- and 8-mers)
in the former.   The 
6-mer \dist\ of the real genomes looks more like the \dist\ of 
a genome that grew randomly one base at a time only to 10 kb long, 
but not to one (or several) Mb long.   We have shown that 
the full-length microbial genome could have grown randomly and 
have such an unexpected \kmer\ \dist\ provided that it grew mostly 
by random self-copying.   And we propose that it is this 
stochastic process, instead of some unknown biological process, 
that has caused the long genome to retain the statistical 
characteristics of its much shorter ancient self. 

Because the probability that a random stretch of DNA would be a 
gene (that codes an RNA or a protein that would fold and function) 
is so minuscule, a population of genomes that stumbled upon a 
self-copying mechanism 
would have had an enormous evolutionary advantage over another 
unfortunate population that did not. 
The preponderance of intragenomic and intergenomic homologous genes 
\cite{Li97} across all life forms 
is testament to the importance of this mechanism 
\cite{ecol,mjan,IHGSC,Celera,bsub}.  

Self-copying growth may not be the only mechanism through which
microbial genomes acquire the statistical characteristics of a 
much shorter sequence.  Such characteristics may 
well have an as yet unknown biological rather than stochastic origin. 
Our model has the virtue of simplicity.  It also has several interesting 
implications of which two are mentioned here. 
The model predicates the coexistence of the two competing modes 
of evolution: the gradual changes of 
classical Darwinism and the stochastic spurts as 
envisioned  in ``punctuated equilibrium'' \cite{Gould,Bak}.   
The fact that a present-day long genome shares vital characteristics 
of its theoretical shorter earlier self implies one knows something 
about its {\it ancestor}, or the 
common ancestor of its relatives.  Perhaps, by pushing this notion harder 
and examining the 
genomes closer, one may gain a deeper understanding of our 
universal ancestor \cite{Woese97}.  

This work is partially supported by a National Science Council 
grant NSC 90-2119-M-008-019.

%\vspace{-10pt}

\end{document}